\documentstyle[sprocl]{article}
\def\Journal#1#2#3#4{{#1} {\bf #2}, #3 (#4)}

\def\PRL{\em Phys. Rev. Lett.}
\def\PRD{{\em Phys. Rev.} D}

\def\mco{\multicolumn}

\def\be{\begin{equation}}
\def\ee{\end{equation}}
\def\bea{\begin{eqnarray}}
\def\eea{\end{eqnarray}}

\begin{document}
\title{TECHNICOLOR MECHANISMS FOR SINGLE TOP PRODUCTION }
\author{LESLEY L. SMITH }
\address{Butler County Community College, 901 S. Haverhill Rd.,
El Dorado, KS 67042}
\author{PANKAJ JAIN }
\address{Indian Institute of Technology, Kanpur, 208016, UP, India}
\author{ DOUGLAS W. MCKAY }
\address{ Department of Physics and Astronomy, University of Kansas, 
\\ Lawrence, KS 66045 }
\maketitle\abstracts{
We investigate the contribution of technicolor mechanisms to the production of 
single top quarks at hadron colliders.  We find that a promising candidate 
process is gluon-gluon fusion to produce a W-boson plus technipion, with 
subsequent decay of the technipion to a top quark plus a bottom
quark.  The top-plus-bottom mode is the dominant one when the technipion mass
is larger than the top mass.  We calculate the total cross section and the 
$p_{T}$ distribution for the technipion production at Tevatron and LHC 
energies for a range of technipion masses, starting at 200 GeV.   The decay 
chain of technipion to top plus bottom quarks and then top to W plus bottom 
yields a final state with two W's and two bottom quarks.  We study the 
backgrounds to our process and the kinematic cuts that maximize the signal
to background.  
We report event rate estimates for the upgraded Tevatron
and the LHC.  }
The discovery of the top quark \cite{CDF} has opened up 
an exciting area of physics.  Top production by non Standard Model (SM)
mechanisms are of great value in probing new physics.
It has been known for some time that the SM
is not completely satisfying.  It does not explain electroweak symmetry breaking
 and has many free parameters.
Technicolor (TC) has emerged as one of the possible successors to the SM,
and we present results of single top production by TC here.
We know the low-energy region of QCD can be represented by a nonlinear
sigma-type model \cite{Peskin}.  Thus we invoke a low-energy effective
Langrangian for a class of technicolor theories which is motivated by
the chiral symmetry of the Technicolor Lagrangian and the observation 
of vector-meson dominance in QCD \cite{Peskin}.
We are interested in the production of single top quarks and we look at the 
gluon fusion process $g g \rightarrow W + P_{8}$
where $P_{8}$ indicates a color octet, SU(2) doublet technipion \cite{example}.
The technipion predominantly decays to a top quark and a bottom quark, 
assuming $M_{P_{8}} > m_{t}$ \cite{EHLQ}.  
The lowest order processes contributing to 
this gluon fusion
are: the $ggWP_{8}$ 4-point interaction diagram, the s-channel 
gluon exchange diagram, and the 2 diagrams with $P_{8}$ exchange in the t 
and u channels.  In the Farhi-Susskind type model that we use for the
present analysis, the vertices needed are the single pseudoscalar effective
Lagrangian
${\cal L}(\phi)_{eff}   =
\frac{iN_{TC}}{16 \pi^{2} F_{T}} \epsilon^{\mu \nu \lambda \rho} \int dx
(\partial_{\mu} G^{b}_{\nu} + \frac{1}{2} g_{3} f^{bcd} G^{c}_{\mu}G^{d}_{\nu})
g_{2}g_{3}(P_{8}^{+,b} \partial_{\lambda} W^{-}_{\rho} + P_{8}^{-,b}
\partial_{\lambda} W^{+}_{\rho})$, 
the normal parity $P_{8}$-$P_{8}$-G effective interaction Lagrangian
\cite{Mckay86} and the usual QCD triple gluon vertex. 
Using these vertices and the tree-level diagrams described above, we 
computed the order $\alpha_{2}^{2} \alpha_{3}$ $p_{T}$ distributions and the
total cross section for technipion production at Tevatron and LHC energies.
The final state that results from the $P_{8}+W$ production and 
subsequent decay of $P_{8}$ into $t + b$ and then $t$ into $W+b$ is
the same $WWbb$ final state that results from $t+ \overline{t}$ production
 and decay.  We comment below on cuts that reduce the $t + \overline{t}$
background.
The $p_{T}$ distribution for $p \overline{p} \rightarrow W + P_{8}$ 
has its peak at $p_{T} \simeq$ 0.1 TeV for $M_{P_{8}}=240$ GeV
and $\sqrt{s}$=14 TeV.  We show several $\frac{d \sigma}{dy dp_{T}}$ values
in the range between 0.1 TeV and 1 TeV for $y=0$, $M_{P_{8}}=240$ GeV,
and $N_{TC}=3$ in Table 1 in units of $fb$ TeV$^{-1}$.
\begin{table}\caption{ $p_{T}$ distribution}
\vspace{0.4cm}
\begin{center}
\begin{tabular}{|l|c|c|c|c|c|c|} \hline
$p_{T}$ (TeV) &0.1  & 0.2  & 0.4  & 0.6  & 0.8  & 1.0  \\
\hline
$\frac{d \sigma_{TC}}{dy dp_{T}}$ ($fb$ TeV$^{-1}$) &
62.0  & 46.6 & 14.4 & 
4.3 & 1.4 & 0.5 \\ \hline	
\end{tabular}
\end{center}
\end{table}
In Table 2 we give the corresponding total cross sections for 
technipion production for a range of technipion masses.  We also show the
events per year expected with the upgraded Tevatron and with the LHC at 
luminosities of $2 \times 10^{32} cm^{-2} s^{-1}$ and $10^{34} cm^{-2} s^{-1}$,
respectively.  At Tevatron energies and expected luminosities, the event rate 
is unobservably small.  At the LHC, however, the event rate is large enough 
to be interesting, allowing rather severe cuts to be made to reduce background
and not lose the whole signal. 
\begin{table}\caption{Technipion cross sections and events per year}
\vspace{0.4cm}
\begin{center}
\begin{tabular}{|l|c|c|c|c|} \hline
  & $M_{P_{8}}=$  
240 GeV & 300 GeV & 350 GeV & 400 GeV \\ \hline
  \mco{5}{|c|}{$\sigma_{TC}$} \\ \hline
$\sqrt{s}=$ 2 TeV & 0.059 fb  & 0.016 fb  & 0.0059 fb 
&  0.0022 fb  \\
$\sqrt{s}=$14 TeV  & 56 fb & 31 fb & 20 fb & 14 fb \\ \hline
  \mco{5}{|c|}{ \# events per year } \\ \hline
$\sqrt{s}=$ 2 TeV & 0.37 & 0.10 & 0.04 & 0.01 \\ \hline
$\sqrt{s}=$ 14 TeV & $1.7 \times 10^{4}$ & $ 9.7 \times 10^{3}$ &
$6.3 \times 10^{3}$ & $4.4 \times 10^{3}$ \\ \hline
\end{tabular}
\end{center}
\end{table}
For $\sqrt{s}$=14 TeV, and a top mass of 180 GeV, the single top cross
section is 117 pb, while the $t \overline{t}$ cross section is 525 pb
\cite{Baringer}.  
Monte Carlo studies indicate that the $WWbb$ decay products of the
TC signal differ from those of the 
SM $t \overline{t}$ production and decay in that the TC $p_{T}$ spectrum
of the recoil $W$ is harder than the $W$ spectrum of $t \overline{t}$ and
the  $b$ and $\overline{b}$ are not back-to-back.   
For example if we impose a 
typical $p_{T} >$ 400 GeV cut on $W$ and a cos $\theta>$0.6 on
 on the $b \overline{b}$ opening angle, 
the fraction of SM $t \overline{t}$ events 
passing these cuts is less than $10^{-3}$ while the fraction of TC events 
passing the cuts is about $5 \times 10^{-2}$.  By tuning these cuts, one   
can substantially reduce the SM $t \overline{t}$ background
in comparison to the TC signal.  Thus our preliminary study indicates that 
a search for charged, colored technipions in the $WWbb$ final state at the LHC
is feasible.

\section*{Acknowledgements}
We thank Phil Baringer for discussions and for providing Monte Carlo studies
of $t$ and $ \overline{t}$ production.  This research was supported in part
by DOE grant \# DE - FG02 - 85ER40214.  The computational facilities of the
Kansas Institute for Theoretical and Computational Science were used 
for part of this work.  D.W.M. thanks S.Ranjbar-Daemi and Faheem Hussain for
the hospitality of the high energy group 
at ICPT, Trieste, during the course of this work.


\begin{thebibliography}{99}
\bibitem{example} For example, in comparison to the gluon fusion process,
direct production of $P_{8}$ from $ t b \overline{b}$ in the sea is
down by more than an order of magnitude.
\end{thebibliography}
\end{document}